# The effect of internal pressure on the tetragonal to monoclinic structural phase transition in ReOFeAs: the case of NdOFeAs

Michela Fratini[1], Rocchina Caivano[1], Alessandro Puri[1], Alessandro Ricci[1], Zhi-An Ren[2], Xiao-Li Dong[2], Jie Yang[2], Wei Lu[2], Zhong-Xian Zhao[2], Luisa Barba[3], Gianmichele. Arrighetti[3], Maurizio Polentarutti[4], Antonio Bianconi[1]

[1]Dipartimento di Fisica, Università di Roma "La Sapienza", P. le Aldo Moro 2, 00185 Roma, Italy; [2]National Laboratory for Superconductivity, Institute of Physics and Beijing National Laboratory for Condensed Matter Physics, Chinese Academy of Sciences, P. O. Box 603, Beijing 100190, P. R. China; [3]Institute of Crystallography, National Council of Research, Elettra, 34012 Trieste, Italy; [4]Structural Biology Lab., XRD1 Diffraction Beamline, Sincrotrone Trieste, Area Science Park, 34012 Trieste. Italy

E-mail:antonio.bianconi@roma1.infn.it and michela.fratini@roma1.infn.it

**Abstract.** We report the temperature dependent x-ray powder diffraction of the quaternary compound NdOFeAs (also called NdFeAsO ) in the range between 300 K and 95 K. We have detected the structural phase transition from the tetragonal phase, with *P4/nmm* space group, to the orthorhombic or monoclinic phase, with *Cmma or P112/a1 (or P2/c)* space group, over a broad temperature range from 150 K to 120 K, centered at $T_0$ ~137 K. Therefore the temperature of this structural phase transition is strongly reduced, by about ~30K, by increasing the internal chemical pressure going from LaOFeAs to NdOFeAs. In contrast the superconducting critical temperature increases from 27 K to 51 K going from LaOFeAs to NdOFeAs doped samples. This result shows that the normal striped orthorhombic *Cmma* phase competes with the superconducting tetragonal phase. Therefore by controlling the internal chemical pressure in new materials it should be possible to push toward zero the critical temperature $T_0$ of the structural phase transition, giving the striped phase, in order to get superconductors with higher $T_c$.

Understanding the quantum mechanism that allows a macroscopic quantum condensate, a superfluid or a superconductor, to resist the decoherence effects of high temperature is a major topic in condensed matter, quantum computing and in the search for quantum mechanisms in the living cell. The evidence that in nature it is possible to achieve a quantum condensate of fermions at high temperatures is provided by the so called high $T_c$ superconductors. Recent experimental results in this field have added to the known high $T_c$ superconductors a new class of materials: the layered pnictide-oxide quaternary compounds ReOTmPn (Re=La, Nd, Ce, Sm …; Tm=Mn, Fe, Co, Ni ; Pn=P, As) [1-12] where the chemical potential can be driven to a particular point of their electronic phase diagram by controlling the charge density [5], the pressure [6], the internal chemical pressure [12], and the lattice disorder, where a superconducting phase with $T_c$ as high as 55 K shows up. The electronic and magnetic properties have been investigated [13-16] and a particular attention has been addressed to the structural properties of the parent compound LaOFeAs. There have been two experimental reports on the low temperature structure of LaOFeAs [17, 18]. The neutron powder diffraction [17] and the x-ray diffraction [18] experiments have demonstrated that LaOFeAs undergoes an abrupt structural distortion below 165 K, changing the symmetry from the tetragonal (space group *P4/nmm*) to the lower symmetry space group (*Cmma* or *P112/n* or *P2/c*) at low temperature that has been shown to be actually monoclinic. In the neutron study [17] it was also reported that the system develops an itinerant antiferromagnetic order with a small magnetic moment (long range spin density wave (SDW)) with a simple stripe like magnetic structure below the magnetic transition temperature at 130 K. This transition does not show up in superconducting doped samples, therefore it has been suggested that the high $T_c$ superconducting phase occurs in proximity of a magnetic stripes ordered phase [19-24].

These results show that the chemical potential in the stoichiometric parent material is already well tuned to the phase that shows the static magnetic stripes order (like the static stripes phase at 1/8 in Nd doped La2124 samples) and by tuning the chemical potential in the vicinity of this quantum critical point high $T_c$ superconductivity appears.

It is therefore of high interest to investigate the parent stoichiometric compounds of these new high Tc superconductors to study how this stripes phase is dependent on chemical pressure. The recent report on the stoichiometric compounds LaFePO has shown that it is not superconductor and it does not show the tetragonal to orthorhombic phase transition [25].

In this work we have investigated the stoichiometric system NdOFeAs, in which the FeAs metallic layer suffers a larger compressive microstrain due to the lattice mismatch with the ReO spacer layers in comparison with LaOFeAs system. Since the chemical pressure controls the superconducting critical temperature $T_c$ in the doped ReOFeAs systems like in cuprate systems [26-28] it is relevant to know the response of the critical temperature of the structural tetragonal to orthorhombic phase transition on the chemical pressure.

The NdOFeAs powder samples were synthesized in Bejing as described elsewhere [8]. The sample consisted of powder of NdOFeAs kept in a 0.5 mm capillary. The temperature of the sample was varied by means of a 700 series Oxford Cryosystems cryocooler with an accuracy of ~2 K. The X-ray diffraction patterns were recorded at the X-ray Diffraction beam-line at the Elettra synchrotron radiation facility in Trieste. The X-ray beam emitted by the wiggler source on the Elettra 2 GeV electron storage ring was monochromatized by a Si(111) double crystal monochromator, focused on the sample with a photon energy of 13.76010 KeV (wavelength 0.90105Å), using a 345 mm ImagePlate-based X-ray detector system (Mar345) (Marresearch GmbH, Norderstedt, Germany). As the intensity of the synchrotron radiation beam slowly decreased during the experiment, every diffraction pattern was collected under the condition that the dose of photons scattered by the sample was constant. The two-dimensional patterns collected with the Mar345 were calibrated with silicon crystalline powder and integrated using the software FIT2D. The rietveld analysis of the XRD powder diffraction patterns was carried out by the GSAS program.

In the range between 300 K and 200 K the XRD diffraction pattern of the NdOFeAs shows the typical tetragonal structure with *P4/nmm* space group. The profile of the 220 line is shown in Figure 1. By decreasing the temperature below 200 K to 150 K the line shows an increasing broadening that increases rapidly in the range between 150 K and 137 K. The data are fitted with the *Cmma* space group below 137 K where we can clearly see the splitting of this line into the two lines of the *Cmma* space group that are indexed as 040 and 400. The phase transition from tetragonal to orthorhombic phase can be followed by plotting the ratio between the intensity of the XRD diffraction signal at $2\theta_1=37.51$ (the center of the 400 line of the *Cmma* space group) and the intensity at $2\theta_2=37.43$ (the center of the 220 line of the *P4/nmm* space group). The results plotted in Fig. 2 show that the tetragonal to orthomibic phase transition, centered at 137 K , is a 30 K wide transition extending from 150 K to 120 K and the 040 and 400 lines are well resolved with our experimental resolution only below 137 K.

In Fig. 3 we report the a and b lattice parameters of NdOFeAs as a function of temperature obtained by GSAS fitting. The present results are compared with the data for LaOFeAs reported by Nomura et al. [18]. Clearly the chemical compressive internal pressure due to lattice mismatch between FeAs and ReO layers in NdOFeAs is larger than in LaOFeAs as it is indicated by the shorter Fe-Fe distance.

Both systems show a similar tetragonal to orthorhombic phase transition at low temperature. The amplitude of the orthorhombic distortion probed by the difference between the a and b axis in the *Cmma* phase is similar in the two samples. The main effect of the larger chemical pressure appears to be the decrease of the critical temperature of the structural phase transition.

In conclusion we have observed the structural phase transition centered at 137 K in NdOFeAs that is very similar to the phase transition at 165 K in LaOFeAs [18]. Therefore there is a clear evidence that the critical temperature for the structural phase transition associated with the onset of a stripes phase, is about 30 K lower that in LaOFeAs. This structural phase transition is similar to the LTO to LTT phase transition in the case of doped cuprates with doping 1/8 and chemical pressure larger than a critical value, that has been identified as the value where two times the $CuO_2$ microstrain exceeds 8% [26-27], associated with the onset of the magnetic stripes order that competes with superconductivity [28,29,30]. Therefore we propose that the superconducting doped layered pnictide-oxide quaternary compounds are similar to cuprates in the overdoped phase and the parent undoped systems are similar to the cuprates at doping 1/8 and large microstrain that show the stripes phase [28-30].

The physical details of this phase transition and the possible phase separation, with the role of the disorder, will deserve further experimental work. This result suggests that the high $T_c$ superconducting phase appears near a quantum critical point where the striped orthorhombic *Cmma* phase competes with the superconductive tetragonal phase. Therefore by changing the internal chemical pressure it should be possible to push toward zero the critical temperature of the stripes phase in order to get superconductor with higher $T_c$.

**Acknowledgments:** We thank for help and discussions Nicola Poccia, Valerio Palmisano, Alessandra Vittorini-Orgeas, and Naurang L. Saini. We acknowledge financial support from European STREP project 517039 "Controlling Mesoscopic Phase Separation" (COMEPHS) (2005).

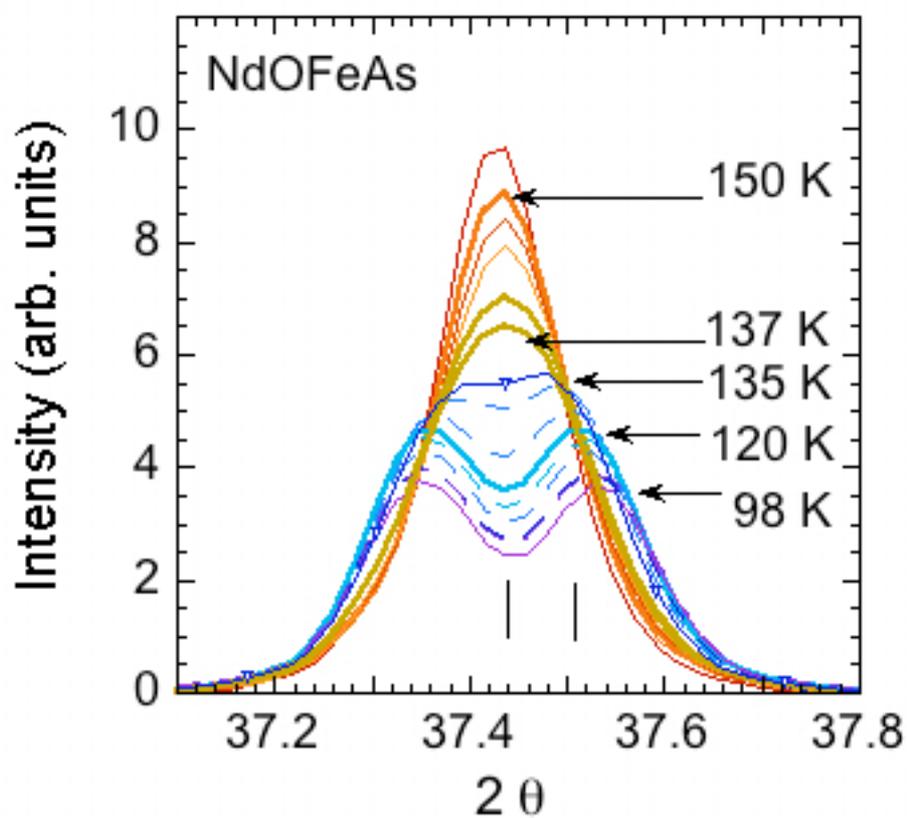

**Fig. 1** The profile of the 220 XRD diffraction line of the tetragonal structure with *P4/nmm* space group in the high temperature phase that is transformed below 137 K into the 040 and 400 lines of the orthorhombic phase with *Cmma* space group.

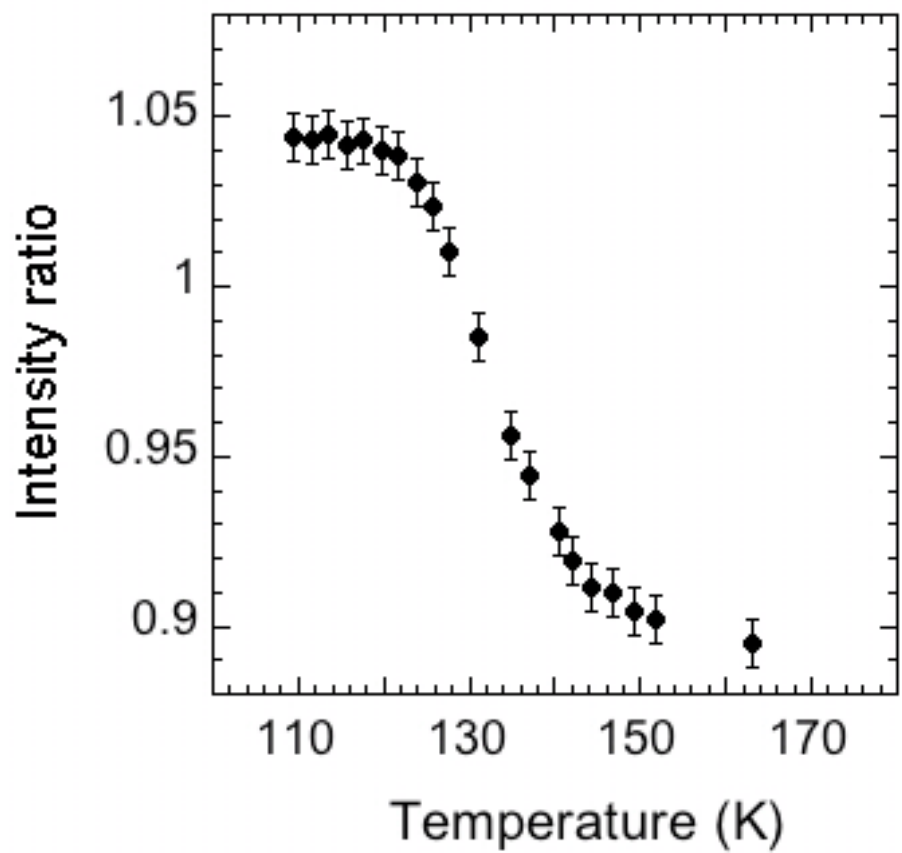

**Fig. 2**. The ratio of the intensity of the XRD diffraction signal at $2\theta_1=37.51$ (the center of the 400 line of the *Cmma* space group) $2\theta_2=37.43$ (the center of the 220 line of the *P4/nmm* space group) probing the broad tetragonal to orthorhombic transition extending from 150K to 120 K, the 040 and the 400 are well resolved with our experimental resolution below the critical temperature that is close to 137 K.

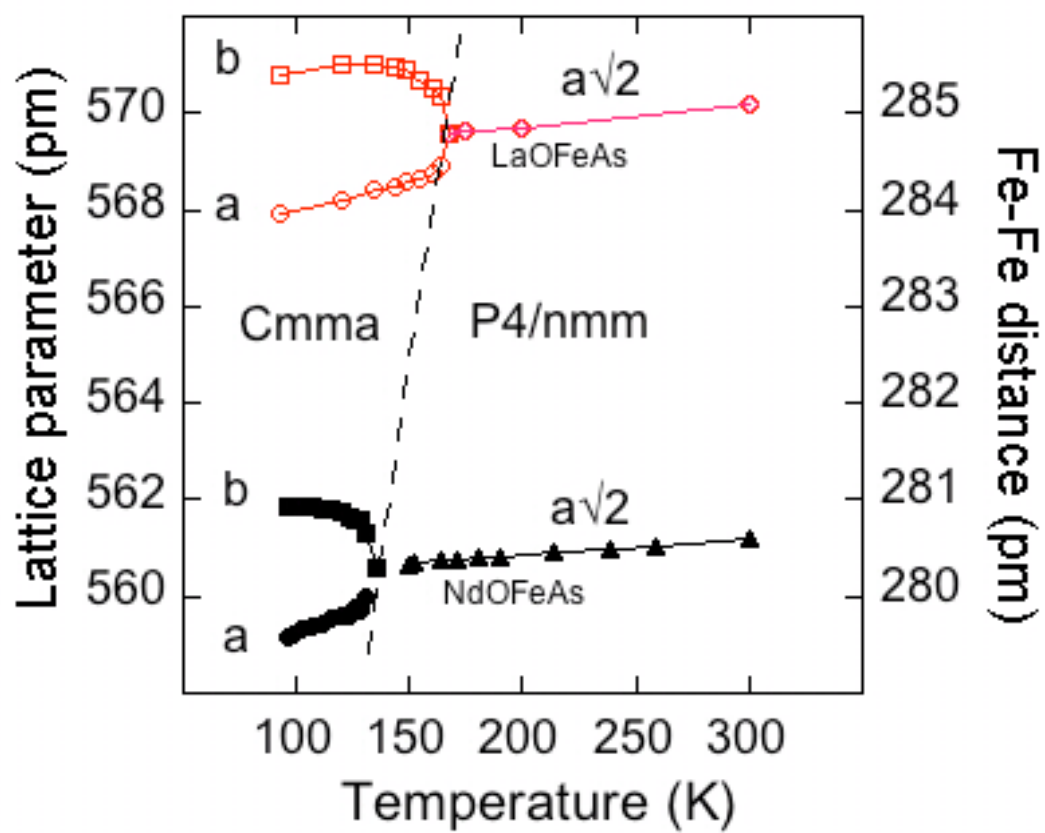

**Fig. 3** The lattice parameters a and b of NdOFeAs as a function of temperature extracted by GSAS are compared with the lattice parameters of LaOFeAs reported in ref. 18.